\documentclass{MolSpaLab}
\usepackage{graphicx}

\newcommand{\emm}[1]{\ensuremath{#1}}   
\newcommand{\emr}[1]{\emm{\mathrm{#1}}} 
\newcommand{\unit}[1]{\emm{\, \emr{#1}}}

\newcommand{\K}   {\unit{K}}
\newcommand{\Kpccm}{\unit{K\,cm^{-3}}}
\newcommand{\pscm}{\unit{cm^{-2}}}
\newcommand{\pccm}{\unit{cm^{-3}}}
\newcommand{\kms}{\unit{km\,s^{-1}}}
\newcommand{\mm}  {\unit{mm}}
\newcommand{\pc}    {\unit{pc}}
\renewcommand{\deg}{\emm{^\circ}}

\newcommand{\e}[1]{\emm{\;10^{#1}}}

\newcommand{\eg} {{\em e.g.}}

\newcommand{\HH}   {\mbox{H$_2$}}       
\newcommand{\CeiO} {\mbox{C$^{18}$O}}   
\newcommand{\CCH}{\mbox{CCH}}
\newcommand{\CCCCH}{\mbox{C$_4$H}}
\newcommand{\cCCCHH}{\mbox{c-C$_3$H$_2$}}
\newcommand{\DCOp}  {\mbox{DCO$^{+}$}}       
\newcommand{\HCOp}  {\mbox{HCO$^{+}$}}       
\newcommand{\HthCOp}{\mbox{H$^{13}$CO$^{+}$}}

\newcommand{\CS}  {\mbox{CS}}        
\newcommand{\CtfS}{\mbox{C$^{34}$S}} 
\newcommand{\HCSp}{\mbox{HCS$^{+}$}} 

\newcommand{\FigPDR}{%
  \begin{figure}
    \centering
    \includegraphics[height=0.9\textwidth{},angle=270]{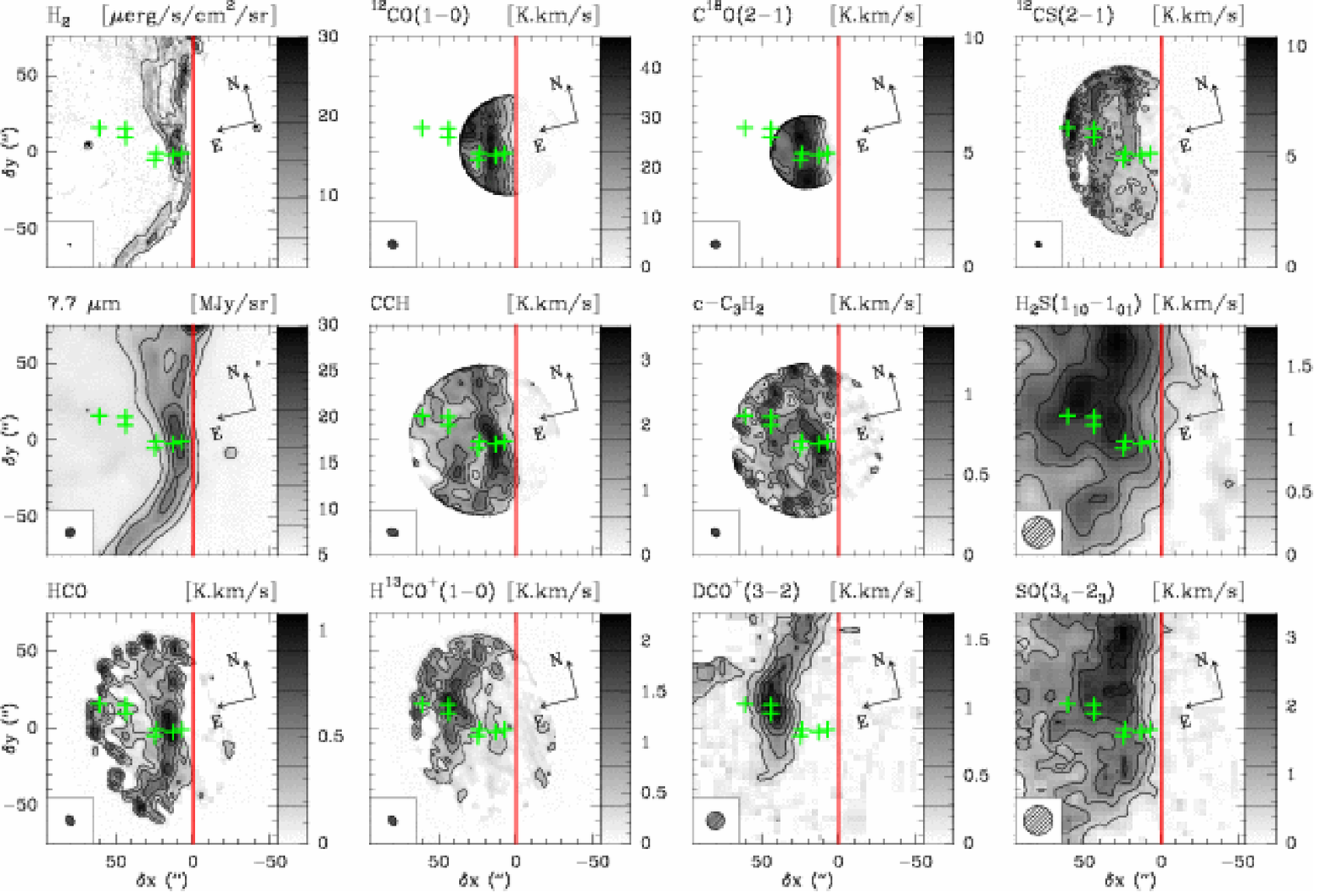}
    \caption{Emission maps obtained with the IRAM Plateau de Bure 
      Interferometer or 30m single-dish, except for the \HH{} v=1-0 S(1)
      emission observed with the NTT/SOFI and the PAH mid--IR emission
      observed with ISO-LW2. The maps have been rotated by 14\deg{}
      counter--clockwise around the image center to bring the exciting star
      direction in the horizontal direction as this eases the comparison of
      the PDR tracer stratifications. Maps have also been horizontally
      shifted by $20''$ (compared to the projection center: RA =
      05h40m54.27s, Dec = -02\deg{}28'00$''$, J2000) to set the horizontal
      zero at the PDR edge delineated as the vertical red line. Either the
      synthesized beam or the single dish beam is plotted in the bottom
      left corner. The emission of all the lines is integrated between 10.1
      and 11.1~\kms{}.  Values of contour level are shown on each image
      lookup table (contours of the \HH{} image have been computed on an
      image smoothed to $5''$ resolution). The green crosses display the
      positions where we derived column densities and abundances (see
      Table~\ref{tab:abundances}).}
    \label{fig:pdr}
  \end{figure}}

\newcommand{\TableAbundances}{%
  \begin{table}[htbp]
    \centering
    \small{%
    \begin{tabular}{cccccc}
      \hline
      \hline
      Species & ($\delta$RA,$\delta$Dec) & ($\delta x$,$\delta y$) & Angular Res. & Col. Dens. & Abundance   \\
              &       [arcsec]           &     [arcsec]            &  [arcsec]    & $N(\mbox{X})$ [\pscm{}] & $\frac{n_\emr{X}}{n_\emr{H}+2n_\emr{H_2}}$ \\ 
      \hline                                                
      \HH{}     &      $(-12,-4)$ &   $(+7.4,-1.0)$ &               12 & $(3.6 \pm 1.7)\e{21}$ & 0.5        \\
      \CeiO{}   &                 &                 & $6.5 \times 4.3$ & $(1.0 \pm 0.3)\e{15}$ & 1.4\e{-7}  \\
      \CCH{}    &                 &                 & $7.2 \times 5.0$ & $(1.1 \pm 0.3)\e{14}$ & 1.5\e{-8}  \\
      \cCCCHH{} &                 &                 & $6.1 \times 4.7$ & $(9.5 \pm 5.0)\e{12}$ & 1.3\e{-9}  \\
      \CCCCH{}  &                 &                 & $6.1 \times 4.7$ & $(3.7 \pm 1.0)\e{13}$ & 5.2\e{-9}  \\
      \HH{}     &       $(-6,-4)$ &  $(+12.2,-2.4)$ &               12 & $(1.1 \pm 0.4)\e{22}$ & 0.5        \\
      \CeiO{}   &                 &                 & $6.5 \times 4.3$ & $(4.0 \pm 0.5)\e{15}$ & 1.9\e{-7}  \\
      \CCH{}    &                 &                 & $7.2 \times 5.0$ & $(3.0 \pm 0.5)\e{12}$ & 1.4\e{-8}  \\
      \cCCCHH{} &                 &                 & $6.1 \times 4.7$ & $(2.4 \pm 1.0)\e{13}$ & 1.1\e{-9}  \\
      \CCCCH{}  &                 &                 & $6.1 \times 4.7$ & $(4.0 \pm 1.0)\e{13}$ & 1.9\e{-9}  \\
      \HH{}     &       $(+6,-4)$ &  $(+24.9,-5.3)$ &               12 & $(2.7 \pm 1.0)\e{22}$ & 0.5        \\
      \CeiO{}   &                 &                 & $6.5 \times 4.3$ & $(5.8 \pm 0.5)\e{15}$ & 1.1\e{-7}  \\
      \CCH{}    &                 &                 & $7.2 \times 5.0$ & $(5.5 \pm 1.0)\e{13}$ & 1.0\e{-9}  \\
      \cCCCHH{} &                 &                 & $6.1 \times 4.7$ & $(2.3 \pm 0.7)\e{12}$ & 4.3\e{-11} \\
      \CCCCH{}  &                 &                 & $6.1 \times 4.7$ & $(2.0 \pm 1.0)\e{13}$ & 3.7\e{-10} \\
      \CS{}     &        $(+4,0)$ &  $(+23.9,-1.0)$ &               10 & $(8.1 \pm 1.0)\e{13}$ & 2.5\e{-9}  \\
      \CtfS{}   &                 &                 &               16 & $(3.7 \pm 0.5)\e{12}$ & 1.2\e{-10} \\
      \HCSp{}   &                 &                 &               29 & $(6.8 \pm 0.5)\e{11}$ & 2.0\e{-11} \\
      \hline
      \HthCOp{} & $(+19.7,+20.4)$ & $(+44.0,+15.0)$ & $6.7 \times 4.4$ &     $(0.5 - 1)\e{13}$ & $(0.5-1)\e{-10}$ \\
      \DCOp{}   &                 &                 &               12 &     $(0.5 - 1)\e{13}$ & $(0.5-1)\e{-10}$ \\
      \CS{}     &     $(+21,+15)$ &  $(+44.0,+9.5)$ &               10 & $(1.2 \pm 1.0)\e{14}$ & 3.9\e{-9}  \\
      \CtfS{}   &                 &                 &               16 & $(5.3 \pm 0.5)\e{12}$ & 1.8\e{-10} \\
      \HCSp{}   &                 &                 &               29 & $(6.8 \pm 0.5)\e{11}$ & 2.3\e{-11} \\
      \CS{}     &     $(+36,+25)$ & $(+61.0,+15.5)$ &               10 & $(1.7 \pm 1.0)\e{14}$ & 6.0\e{-9}  \\
      \CtfS{}   &                 &                 &               16 & $(7.9 \pm 0.5)\e{12}$ & 2.5\e{-10} \\
      \HCSp{}   &                 &                 &               29 & $(9.0 \pm 0.5)\e{11}$ & 3.8\e{-11} \\
      \hline
    \end{tabular}}
    \caption{Column densities and abundances of several chemical species
      from the UV--illuminated PDR (upper part) to the shielded region
      (lower part) of the Horsehead edge. Coordinates offsets are given 
      both in the Equatorial system from RA = 05h40m54.27s, Dec = 
      -02\deg{}28'00$''$ (J2000) and in the coordinate system adapted to 
      the source geometry and used in Figure~\ref{fig:pdr}. Abundances are 
      computed with respect to the density of protons $(n_\emr{H})$.}
    \label{tab:abundances}
  \end{table}}

\begin{document}
\TitreGlobal{Molecules in Space \& Laboratory}
\title{The Horsehead mane: Towards \\
  an observational benchmark for chemical models}
\author{\FirstName J. \LastName Pety}
\address{IRAM \& Observatoire de Paris, France}
\author{\FirstName J. R. \LastName Goicoechea}
\address{LERMA-LRA, France}
\author{\FirstName M. \LastName Gerin$^2$}
\author{\FirstName P. \LastName Hily-Blant}
\address{IRAM \& LAOG, France}
\author{\FirstName D. \LastName Teyssier}
\address{European Space Astronomy Centre, Spain}
\author{\FirstName E. \LastName Roueff}
\address{LUTH, France}
\author{\FirstName E. \LastName Habart}
\address{IAS, France}
\author{\FirstName A. \LastName Abergel$^5$}

\runningtitle{The Horsehead mane as an observational benchmark}
\setcounter{page}{1}

\maketitle
%
%

\vspace*{-0.25cm}
\section{The intrinsic complexity of chemical models}
\vspace*{-0.25cm}

Photodissociation region (PDR) models are used to understand the evolution
of the FUV illuminated matter both in our Galaxy and in external galaxies.
To prepare for the unprecedented spatial and spectroscopic capabilities of
ALMA and Herschel, two different kinds of progresses are currently taking
place in the field.  First, numerical models describing the chemistry of a
molecular cloud are being benchmarked against each other to ensure that all
models agree not only qualitatively but also quantitatively on at least
simple cases\footnote{See also
  \texttt{http://www.ph1.uni-koeln.de/pdr-comparison/intro1.htm}} (R\"ollig
et al.\ 2007).  Second, new or improved chemical rates are being
calculated/measured by several theoretical and experimental groups.
However, the difficulty of this last effort implies that only a few
reactions (among the thousand ones used in chemical networks) can be
thoroughly studied. New numerical tools are thus being developed for taking
into account the impact of the uncertainties of the chemical rates on the
chemical model predictions, and their comparison with observed abundances
(\eg{} Wakelam et al.\ 2005, 2006). In view of the intrinsic complexity of
building reliable chemical networks and models, there is an obvious need of
well-defined observations that can serve as basic references. PDRs are
particularly well suited to serve as references because they make the link
between diffuse and dark clouds, thus enabling to probe a large variety of
physical and chemical processes.

\vspace*{-0.25cm}
\section{The Horsehead edge as a chemical laboratory}
\vspace*{-0.25cm}

\FigPDR{} %
\TableAbundances{} %

The illuminated edge (PDR) of the western condensation presents one of the
sharpest infrared filament (width: $10''$ or 0.02\pc{}) detected in our
Galaxy by ISOCAM. The most straightforward explanation given by Abergel et
al. (2003) is that most of the dense material is within a flat structure
viewed edge-on and illuminated in the plane of the sky by $\sigma$Ori.  The
\HH{} fluorescent emission observed by Habart et al.\ (2005) is even
sharper (width: $5''$), implying the inclination of the PDR on the
plane-of-sky to be less than 5\deg{}. The Horsehead ridge thus offers the
opportunity to study at small linear scales ($1''$ corresponds to
0.002\pc{} at 400\pc{}) the physics and chemistry of a PDR with a simple
geometry, very close to the prototypical kind of source needed to serve as
a reference to chemical models. Since 2001, we started to study the
Horsehead PDR mainly with the IRAM Plateau de Bure interferometer at 3\mm{}
and the IRAM-30m at 1\mm{} achieving spatial resolutions from 3 to $11''$
(except the \HH{} fluorescent emission observed at $1''$-resolution with
the NTT/SOFI instrument). Figure~\ref{fig:pdr} displays all the observed,
high resolution maps already acquired. Those maps trace the different
layered structures predicted by photochemical models according to 
chemical reaction networks, excitation conditions and radiative transfer.

Abergel et al.\ (2003) deduced from the distance between $\sigma$Ori and
the PDR that the intensity of the incident far UV radiation field is $\chi
\sim 60$ relative to the interstellar radiation field in Draine's units.
Through the modelling of the \HH{} and CO emission, Habart et al.\ (2005)
showed that the PDR has a very steep density gradient, rising to $n_\emr{H}
\sim 10^5\pccm$ in less than $10''$ (\ie{} 0.02\pc{}), at a roughly
constant pressure of $P \sim 4 \times 10^6\Kpccm$. These observations were
followed by a chemical study of small hydrocarbons (\CCH{}, \cCCCHH{},
\CCCCH{}).  Pety et al.\ (2005) showed that the abundances of the
hydrocarbons are higher than the predictions based on pure gas phase
chemical models (Le Petit et al.\ 2002 and reference therein). These
results could be explained either by the photoerosion of the large aromatic
molecules and/or small carbon grains (C. Joblin, 2007, this volume) or by a
turbulent mixing which would transport in the illuminated part of the PDR
molecules after their productions in the dark part (M. Gerin, 2007, this
volume). Goicoechea et al.\ (2006) linked the PDR model predictions with
detailed non-LTE, nonlocal excitation and radiative transfer models adapted
to the Horsehead geometry.  They showed that the gas sulfur depletion
invoked to account for CS and HCS$^+$ abundances is orders of magnitude
lower than in previous studies of the sulfur chemistry.

Finally, Pety et al.\ (2007) studied the deuterium fractionation in the
Horsehead edge from observations of several \HthCOp{} and \DCOp{} lines. A
large [\DCOp]/[\HCOp] abundance ratio $(\geq 0.02)$ is inferred at the
\DCOp{} emission peak, a condensation shielded from the illuminating far-UV
radiation field where the gas must be cold (10--20\K{}) and dense
($n_\emr{H} \geq 4 \times 10^5\pccm$). \DCOp{} is not detected in the
warmer photodissociation front, implying a lower [\DCOp]/[\HCOp] ratio ($<
10^{-3}$). To our knowledge, this is the brightest \DCOp{} emission (4\K{})
detected in an interstellar cloud close (angular distance $<40''$) to a
bright \HH{}/PAH emitting region. This opens the interesting possibility to
probe at high resolution the chemical transition from far-UV photodominated
gas to ``dark cloud'' shielded gas in a small field of view.

\vspace*{-0.25cm}
\section{Towards an observational benchmark}
\vspace*{-0.25cm}

An ideal observational benchmark would deliver to chemists a set of
abundances (with the associated uncertainties) as a function of the 
distance (or extinction) to the illuminating star. 
This goal is difficult to achieve for several reasons: 1) The
geometry of the source is never as simple as wished when it is known at
all; 2) The spectra produced by the instruments must be inverted to obtain
abundances; 3) The spectra are often measured at very different angular
resolutions, implying beam dilution and/or mixing of different gas
components. For several years now, we have started to systematically study
the western edge of the Horsehead nebula because its geometry is not only
well understood but also quite simple (almost 1D and viewed edge-on). The
density profile across the PDR is well constrained and there are several
current efforts to constrain the thermal profile. The combination of low
distance to Earth (400\pc{}), low illumination ($\chi \sim 60$) and high
density ($n \sim 10^5\pccm$) implies that all the interesting physical and
chemical processes can be probed in a field-of-view of less than $50''$
with typical spatial scales ranging between 1 and $10''$.

All those observations are done by the same team, using the same
instruments (mainly IRAM Plateau de Bure and 30m) and the same methods both
of data reduction and data analysis. For each species, we are trying to
observe several transitions at similar angular resolutions (from 5 to
$15''$) to constrain properly the excitation conditions and
derive accurate column densities and abundances. Obtaining
emission maps when possible has proved essential to understand the spatial
distributions of the species. We are preparing the public release of cuts
of spectra so that the radiative transfer analysis can be refined as
knowledge of collisional rates progresses. In the meantime,
Table~\ref{tab:abundances} is our first attempt at quantitatively
summarizing the results obtained up to now. In the future, the zoom
capacity and resolving power of ALMA will enable to measure many different
specie transitions at a resolution of $1''$, enabling to resolve all the
physical and chemical gradients.

\end{document}